\documentclass{webofc}

\usepackage[varg]{txfonts}   
\usepackage{hyperref}
\usepackage{url}
\hypersetup{colorlinks=true,citecolor=blue,urlcolor=blue,linkcolor=blue}
\begin{document}
\title{Benchmarking XRootD-HTTPS on 400Gbps Links with Variable Latencies}
%
%

\author{
    \firstname{Aashay} \lastname{Arora}\inst{1}\thanks{\email{aashay.arora@cern.ch}} \and
    \firstname{Diego} \lastname{Davila}\inst{1} \and
    \firstname{Frank} \lastname{W\"urthwein}\inst{1} \and
    \firstname{John} \lastname{Graham}\inst{1} \and
    \firstname{Dima} \lastname{Mishin}\inst{1} \and
    \firstname{Justas} \lastname{Balcas}\inst{2} \and
    \firstname{Tom} \lastname{Lehman}\inst{2} \and
    \firstname{Xi} \lastname{Yang}\inst{2} \and
    \firstname{Chin} \lastname{Guok}\inst{2} \and
    \firstname{Harvey} \lastname{Newman}\inst{3}
}

\institute{
    University of California San Diego / San Diego Supercomputer Center
    \and
    Energy Sciences Network (ESNet)
    \and
    California Institute of Technology
}

\abstract{
In anticipation of the High Luminosity-LHC era, there is a critical need to oversee software readiness for upcoming growth in network traffic for production and user data analysis access. This paper looks into software and hardware required improvements in US-CMS Tier-2 sites to be able to sustain and meet the projected 400 Gbps bandwidth demands while tackling the challenge posed by varying latencies between sites. Specifically, our study focuses on identifying the performance of XRootD HTTP third-party copies across multiple 400 Gbps links and exploring different host and transfer configurations.
Our approach involves systematic testing with variations in the number of origins per cluster and CPU allocations for each origin. By replicating real network conditions and creating network "loops" that traverse multiple switches across the wide area network, we are able to replicate authentic network conditions.
}
\maketitle
\section{Introduction}
\label{intro}
In face of the High Luminosity-LHC (HL-LHC) era coming on in 2030, there is a significant expected gap between the computing requirements and the hardware purchases given the projected budget. In order to make up for such a gap, there are numerous efforts directed into both making software more efficient and identifying possible scalability issues in the current infrastructure. One important part of the infrastructure is the one in charge of transferring files between the different sites. These types of data transfers are commonly referred to as third party copy (TPC) transfers.
TPC transfers are responsible for distributing data across the different institutions that conform scientific collaborations. In the case of the Compact Muon Solenoid (CMS) experiment, millions of these transfers happen every day among 100+ different sites. The infrastructure that supports these type of transfers is composed by the data management system, that orchestrate the transfers, the storage systems that send and receive the data at each site and the network that interconnect them. Making good use of these resources will be imperative in order to achieve the scale of HL-LHC. 
The estimated bandwidth capacity for each of the eight American Tier-2 CMS sites for the HL-LHC era is 400 Gbps\cite{carder2022basic} and while carrying out the necessary network upgrades is a major effort, making sure that the Storage Systems are capable of sustaining the targeted throughput is a completely different challenge. The focus of this work is on XRootD \cite{xrootd,xrootd-paper} which is the software used by all American Tier-2 sites to expose their storage systems.

\section{Background}
\label{background}
Throughput is defined as the measurement of the amount of data transferred between 2 parties (sender and receiver servers) per unit of time, and although this seems like a simplistic metric, there are many other variables in play that affect how much data a sender can send and the receiver can receive. At the server level, we have the system buffers that limit the maximum amount of data that can be in flight at any given time. Then we have the Maximum Transmission Unit (MTU) which dictates the size at which this data has to be chopped into packets. Processing all these packets requires CPU thus the number of cores, available to this systems, also plays an important role in this. Latency, which is the time that it takes to a packet to travel the physical distance between sender and receiver, plays a main role in this equation as there is no way around having every single packet travel this distance. Latency and Round Trip Time (RTT), roughly the double of latency, are commonly (but wrongly) used as interchangeable terms. In this work when we say "latency" we are actually referring to RTT.
Another important variable known as packet loss happens mainly due to network traffic and or faulty equipment and its probability increases with latency, and more importantly how this loss is perceived and act upon by the congestion protocols, also determines how much data the sender is allowed to send at any given time. The number of streams, this is, the number of independent data transfers between sender and receiver, can help increasing the aggregated throughput but not without cost; more streams require more CPU and larger buffers, and might increase the probability of packet loss. Finally, at the end of the stack we have the software that will have to ultimately process all the transferred data. Making sure that the software is able to scale up and deal with the desired throughput is the high-level goal of this work.

\section{Previous Studies}
\label{previous-studies}
In the past we have proven that XRootD can sustain an aggregate throughput of 400 Gbps at an RTT of 5ms\cite{arora2024400gbps},the round trip between the University of California in San Diego (UCSD) and the California Institute of Technology (Caltech). One of the main challenges we faced in the aforementioned experiment was the high number of streams needed to sustain such throughput. Knowing that the distribution of RTT between any pair of Tier-2 sites in the US and from them to CERN, ranges from 5 to 120 ms the next logical step was to verify that XRootD could scale within such range. We carried out a first attempt by inducing artificial latency using the Linux Traffic Control (tc)\cite{tc} and although the resulting trends matched our expectations their magnitude did not. We believe that the reason for this discrepancy is due to the artificial latency and for that reason we decided to conduct this study using real latencies instead.

\section{Testbed Setup}
\label{testbed-setup}
The main objective of this work was to characterize the relationship between throughput, latency, number of streams, CPU and number of XRootD instances when dealing with TPCs. We use Kubernetes\cite{k8s} to manage different configurations of XRootD instances and the CPU cores allocated to them. On the other hand, to have a variety of real RTTs we utilized SENSE\cite{monga2020software} and the FABRIC testbed\cite{fabric} as described in the following sections.

\subsection{Data Transfer Nodes}
\label{testbed-dts}
For our tests we utilized 2 identical servers, sitting next to each other in the San Diego Supercomputer Center (SDSC) as our Data Transfer Nodes (DTNs). They have the following SPECS: 2 x 32-core Intel Xeon Gold 6430, 2 TB of DDR5 RAM and a ConnectX-7 NIC capable of 400Gbps and have been tuned for high throughput over high latency by increasing the maximum read and write buffer sizes to 1 GB via the kernel parameters \textit{net.core.rmem\_max} and \textit{net.core.wmem\_max}. The MTU of both host has been set to 9k.

\subsection{Network Setup}
\label{testbed-network}
\begin{figure}[h]
\centering
\includegraphics[width=10cm,clip]{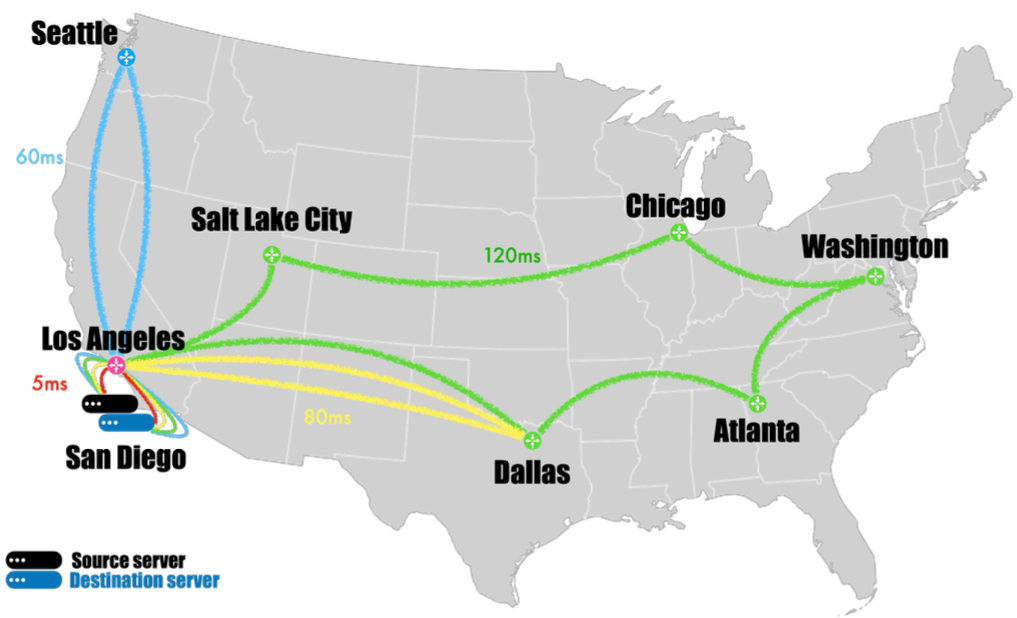}
\caption{Network routes with different latencies interconnecting our DTNs}
\label{map}
\end{figure}

Using SENSE's L2 and routing capabilities, we were able to interconnect our DTNs through a set of different static network routes looping across the FABRIC testbed as shown in Figure~\ref{map}. We picked a range of RTTs between 5 and 120 ms based on the distances among the Tier-2 sites in the US and CERN. In order to avoid possible traffic contention with other experiments going on in FABRIC, we leveraged SENSE's Quality of Service (QoS) feature to request guaranteed bandwidth allocations on each route.  

\subsection{XrootD Deployment}
\label{testbed-network}
We used the Kubernetes cluster of NRP \cite{nrp} to manage the different configurations of CPU cores and number of XRootD instances (or origins) in our tests. We configured our XRootD instances to support TPCs over the HTTP protocol. For tests with more than 1 origin we used the \textit{clustered} configuration of XRootD to balance the load among the origins. In every case, we use a \textit{tmpfs} file system and file sizes of 4 GB each. 

\section{Tests}
\label{tests}
Using a separate Kubernetes pod we ran a bash script that orchestrates a given amount of TPCs by running parallel instances of gfal-copy\cite{gfal} on a separate Kubernetes pod. We designed our tests with the following questions in mind:
\begin{enumerate}
\item What is the effect of increasing latency over throughput and how can we tune the number of streams to attenuate such effect?
\item What is the minimum number of cores needed to reach 100 Gbps?
\item What is the minimum number of cores needed to reach 200 Gbps?
\item What maximum amount of throughput we can get from a single server?
\end{enumerate}

\section{Results}
\label{results}
\begin{figure}[h]
\centering
\includegraphics[width=10cm,clip]{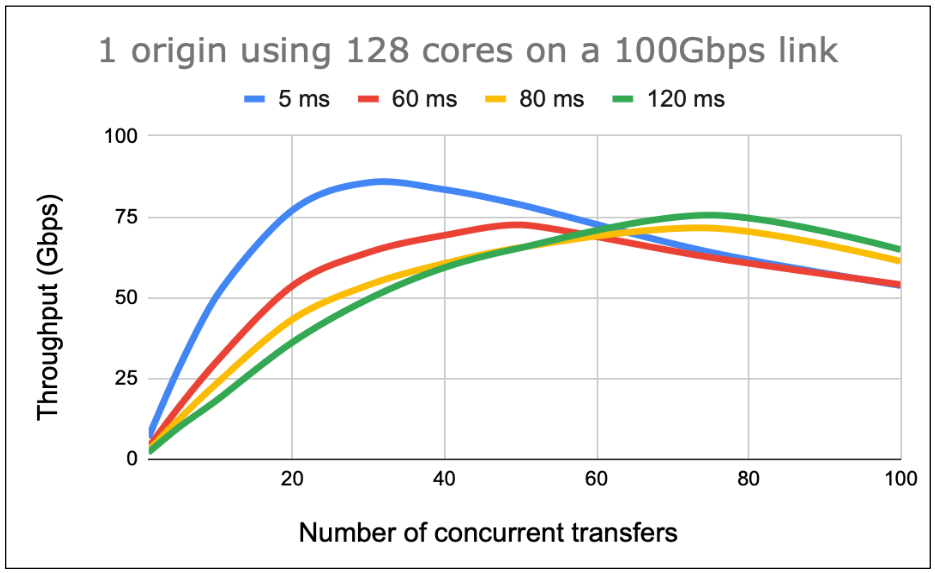}
\caption{Network routes with different latencies interconnecting our DTNs}
\label{1-128-100}
\end{figure}

Initially we had, as rule of thumb, that throughput is inversely proportional to latency and directly proportional to the number of streams, and although this seems to be the common trend, we can see, in figure \ref{1-128-100}, that is not always the case. Another aspect to highlight in this figure is how the distribution followed by the different RTTs is significantly distinct. We can clearly see that for small RTT, throughput climbs rapidly when increasing the number of streams but it also drops considerably fast when a given threshold is passed, and also we can see these effects are softened as latency increases.

\begin{figure}[h]
\centering
\includegraphics[width=10cm,clip]{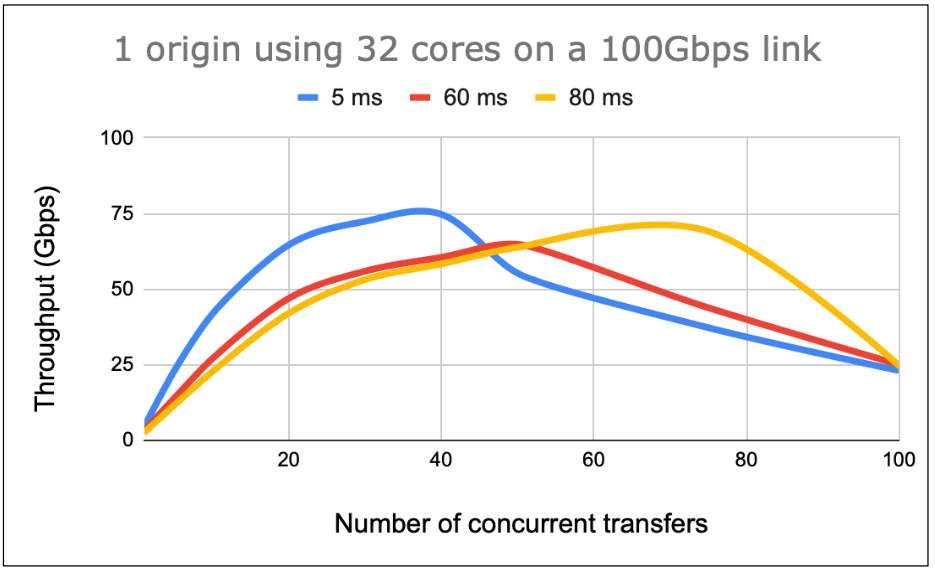}
\caption{Distribution of throughput per number of streams across different RTTs}
\label{1-32-100}
\end{figure}

\begin{figure}%
\centering
\begin{minipage}{.5\textwidth}
  \centering
  \includegraphics[width=6.5cm,clip]{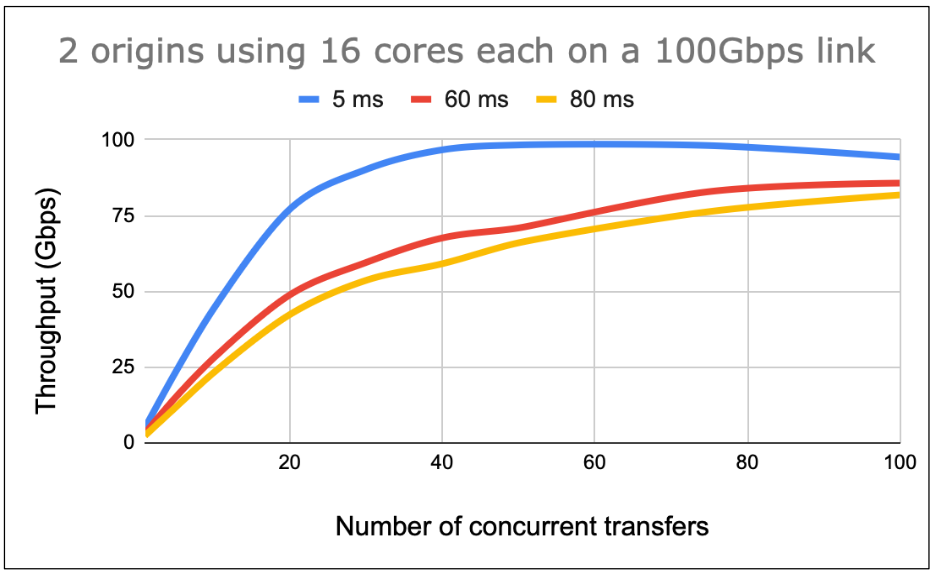}
  \caption{Total cores per server: 32, bandwidth limit: 100 Gbps}
  \label{2-16-100}
\end{minipage}%
\begin{minipage}{.5\textwidth}
  \centering
  \includegraphics[width=6.5cm,clip]{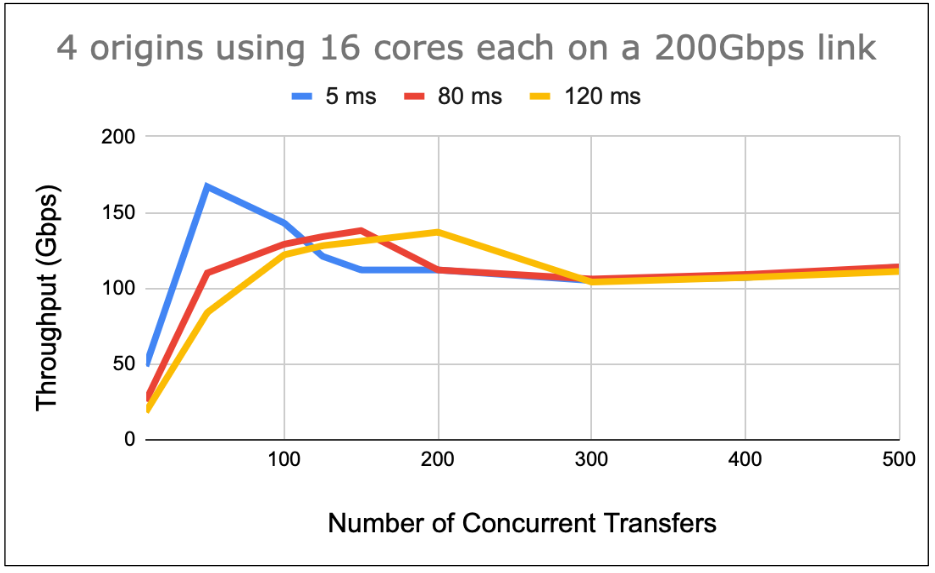}
  \caption{Total cores per server: 64, bandwidth limit: 200 Gbps}
  \label{4-16-200}
\end{minipage}
\end{figure}

Looking at figures \ref{1-32-100}, \ref{2-16-100} and \ref{4-16-200} we can see the evolution of our tests while trying to reach 100 Gbps. One thing to note is that with 1 origin it seems impossible to reach 100 Gbps (figure \ref{1-32-100}). With an additional origin, one can reach 100 Gbps at low latencies by increasing the number of streams, but for large RTTs the target remains unfeasible (figure \ref{2-16-100}). Finally in figure \ref{4-16-200} we can see that with 4 origins and 64 CPU cores in total it is possible, even for large RTTs, to reach 100 Gbps with less than 100 streams.

\begin{figure}%
\centering
\begin{minipage}{.5\textwidth}
  \centering
  \includegraphics[width=6.5cm,clip]{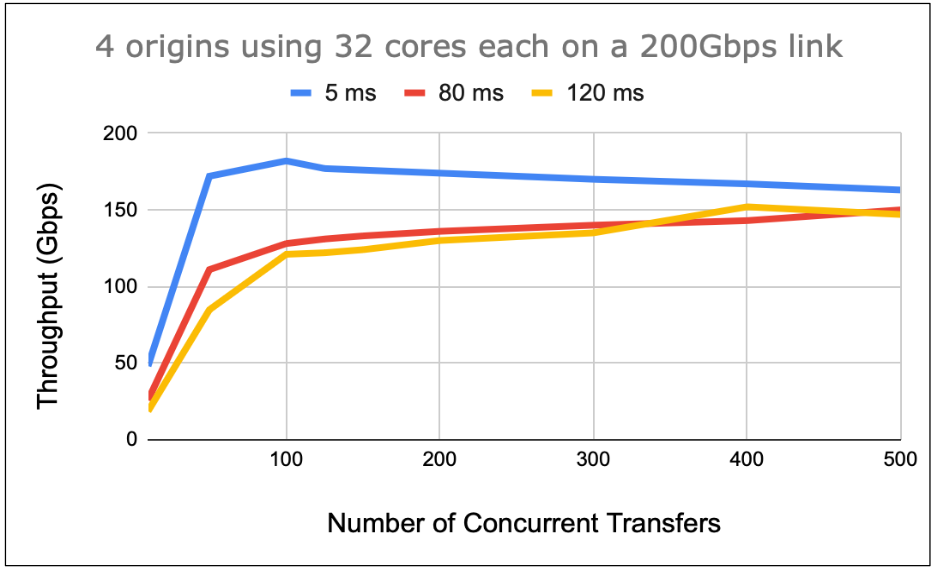}
  \caption{Total cores per server: 128, bandwidth limit: 200 Gbps}
  \label{4-32-200}
\end{minipage}%
\begin{minipage}{.5\textwidth}
  \centering
  \includegraphics[width=6.5cm,clip]{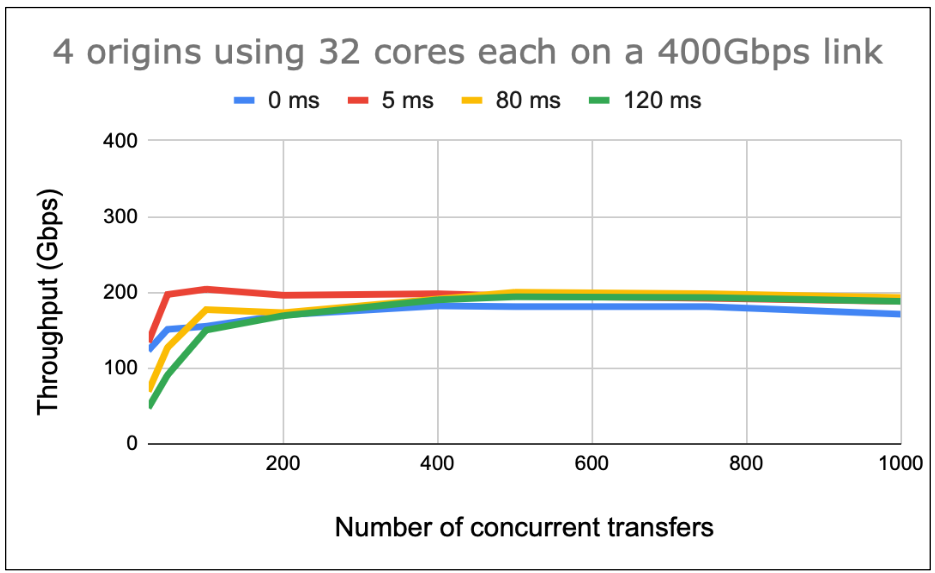}
  \caption{Total cores per server: 128, bandwidth limit: 400 Gbps}
  \label{4-32-400}
\end{minipage}
\end{figure}

Similarly, in the test depicted by figure \ref{4-32-200} we try to achieve 200 Gbps by doubling the number of cores used in figure \ref{4-16-200} but the result makes evident that it takes more than that to reach the target. Interestingly in a similar test, depicted in figure \ref{4-32-400}, where all conditions remain the same except for the bandwidth allocation, which is doubled, we are able to reach the target of 200 Gbps. This indicates a negative effect on throughput when getting close to the bandwidth limit even when the limit is not reached.

\begin{figure}[h]
\centering
\includegraphics[width=10cm,clip]{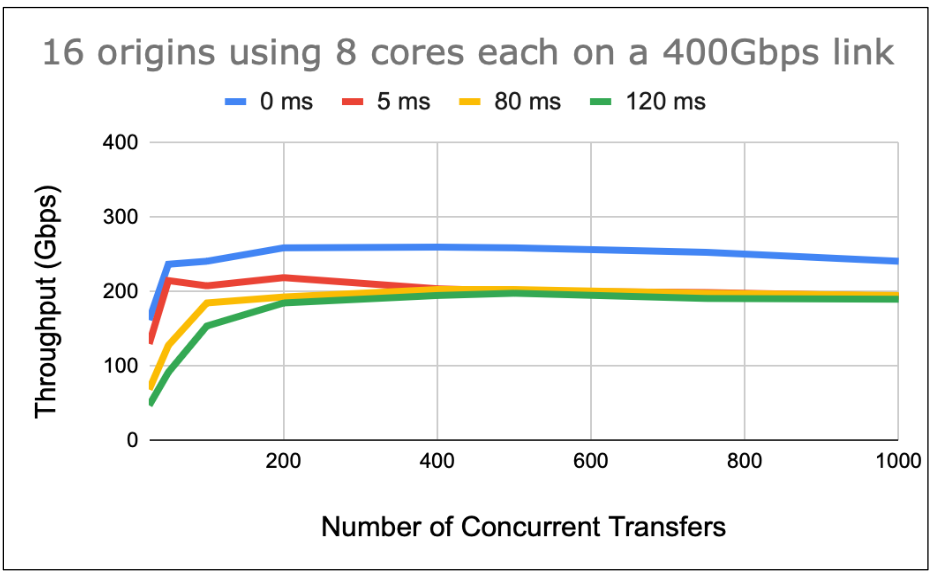}
\caption{Network routes with different RTTs interconnecting our DTNs}
\label{16-8-400}
\end{figure}

Finally, looking at figure \ref{16-8-400} we can tell that there is a hard limitation of about 260 Gbps on a single server even at 0 ms RTT and pushing an excess of hundreds of streams.

\subsection{Other Remarks}
Comparing figures \ref{1-32-100} and \ref{2-16-100} we can see that adding more origins while keeping the total number of cores fixed, makes the overall system perform better. It looks as if a single XRootD instance is not able to scale up past 16-cores.

In figure \ref{2-16-100} we can see how once we saturate the available bandwidth, adding more streams does not have the negative effect on throughput that we see in other figures like \ref{1-128-100}, \ref{4-16-200} and \ref{1-32-100} where throughput is always far from the limit.

\section{Conclusions}
\label{conclusions}
In this study we have shown the effects that latency, number of cores and number of XRootD instances have over throughput in a series of scenarios that mimic TPC transfers in production systems.
Although many of the patterns depicted in our results were expected, we were able to find interesting patterns that could help us tune our systems in order to optimize overall throughput like:

\begin{itemize}
\item Generating high throughput over short latencies requires a lot fewer CPU cores than for longer latencies
\item 4 XRootD origins are needed in order to reach 100 Gbps, comfortably, at long latencies
\item Using XRootD we cannot reach beyond 260 Gbps with a single physical server.
\item Distributing CPU cores among XRootD instances pays serious dividends on throughput
\item The number of streams is a double bladed knife; either too little or too many streams will hurt throughput
\begin{itemize}
    \item The above is less accentuated by short latencies
    \item Once we have reached the bandwidth limit adding more streams does not incurs in penalties
\end{itemize}
\item It is not necessary to reach the bandwidth limit to experience the effects of saturation
\end{itemize}

Finally, we expect this study to serve as a base for improvement for systems like FTS\cite{fts} and DMM\cite{rucio-sense} that try to optimize throughput generated by TPC transfers among many interconnected storage systems.

\section{Acknowledgments}
This work is partially supported by the US National Science Foundation (NSF) Grants OAC-1836650, PHY-2323298, PHY-2121686 and OAC-2112167. Finally, this work would not be possible without the significant contributions of collaborators
at ESNet, Caltech, and SDSC.

\bibliography{references.bib}

\end{document}